\documentclass[pra,twocolumn,aps,showpacs,longbibliography]{revtex4-1}

\usepackage{hyperref}
\hypersetup{
      colorlinks=true,
      citecolor=blue,
      linkcolor=blue,
      urlcolor=blue}
\usepackage{graphicx}
\usepackage{amsmath}
\usepackage{amsfonts}
\usepackage{amssymb}
\usepackage{times}
\usepackage{subfigure}
\usepackage[usenames,dvipsnames]{color}
\usepackage{bm}
\everymath{\displaystyle}
\begin{document}
\title{Interacting bosons in two-dimensional lattices with localized 
dissipation}

\author{Arko Roy}
\author{Kush Saha}
\affiliation{Max-Planck-Institut f{\"u}r Physik komplexer Systeme,
N{\"o}thnitzer Stra\ss e 38, 01187 Dresden, Germany}
\date{\today}


\begin{abstract}
Motivated by the recent experiment [Takafumi Tomita \emph{et al.}, 
Sci. Adv. {\bf 3}, (2017)], we study the dynamics of interacting bosons in a 
two-dimensional optical lattice with local dissipation. 
Together with the Gutzwiller mean-field theory for density matrices and 
Lindblad master equation, we show how the onsite interaction between bosons
affects the particle loss for various strengths of dissipation. For 
moderate dissipation, the trend in particle loss differs significantly near 
the superfluid-Mott boundary than the deep superfluid regime. While the loss 
is suppressed for stronger dissipation in the deep superfluid regime, 
revealing the typical quantum Zeno effect, the loss near the phase boundary 
shows non-monotonic dependence on the dissipation strength. We furthermore 
show that close to the phase boundary, the long-time dynamics is well 
contrasted with the dissipative dynamics deep into the superfluid regime. 
Thus the loss of particle due to dissipation may act as a probe to 
differentiate strongly-correlated superfluid regime from its 
weakly-correlated counterpart.    
	
\end{abstract}



\maketitle

\section{Introduction}\label{Introduction}

Although dissipation in quantum systems leads to decoherence of quantum states,
recent years have witnessed an upsurge in allowing dissipation on purpose to 
study non-equilibrium dynamics in various physical systems such as optical 
cavities~\cite{RevModPhys.85.553}, trapped ions~\cite{Blatt2012,Bohnet1297}, 
exciton-polariton BECs~\cite{RevModPhys.85.299}, and microcavity arrays 
coupled with superconducting qubits~\cite{Houck_2012}. This is in part, 
because dissipation can be used as an efficient tool for preparing and 
manipulating quantum states~\cite{MULLER20121}, and in part because the 
interplay between unitary and dissipative dynamics leads to the emergence
of non-equilibrium steady states~\cite{tomadin_2011}. Among various 
experimental platforms for studying dissipative dynamics, the most
promising candidate turns out to be cold atoms due to its high degree of
experimental controllability. This has led to a recent cold atom experiment
where single and two-body particle losses have been investigated with widely 
controllable dissipation strength, revealing the melting of quantum phases 
across the superfluid-Mott transition~\cite{tomita_17}.

In this work, we consider a two-dimensional Bose-Hubbard model and examine 
the effect of dissipation on the quantum phase transition from a Mott
insulator (MI) to a superfluid phase (SF) regime. In particular, we focus on 
the different parameter regimes of the interaction for a fixed chemical
potential. It is well established that the superfluid phase can itself be 
classified into two regimes depending on the relative strength of the two 
momentum scales, namely the Ginzburg momentum $k_G$ and healing 
momentum $k_h$. For $k_G/k_h\sim1$,  a weakly-correlated superfluid  (WCSF) 
regime is obtained, 
where superfluid stiffness or condensate density reaches to the maximum 
Bogoliubov-approximated values. In contrast, for  $k_G/k_h\ll1$, a 
strongly-correlated superfluid regime (SCSF) emerges, where no Bogoliubov
theory exists and the condensate density and superfluid stiffness are found 
to be highly suppressed~\cite{rancon_11}. In view of that, we aim to address 
the following issues: Is the dynamics different in these two regimes of 
superfluid states? Can we probe these two regimes by studying dissipative 
dynamics? We note that although dissipative dynamics has been studied both in 
few-sites and extended Bose-Hubbard model~\cite{konotop_2013,witthaut_2013,
vidanovic_14,thomas2014}, the effect of dissipation on the different 
superfluid regimes is yet to be explored in detail.  

To study dissipative dynamics of the Bose-Hubbard model, we use the 
Gutzwiller mean-field (GMFT) theory for density matrices. Previous works have 
shown that the GMFT can validate correct dynamics of the Bose-Hubbard model 
in the clean~\cite{kotliar1991,krauth1992} and disordered 
limit~\cite{sheshasdri_1995,buonsante_2007,buonsante_2007a}. Moreover, 
the study of nonequilibrium phenomena such as expansion dynamics of Mott 
phase~\cite{mark_2011}, dynamical generation of molecular 
condensates\cite{jaksch_2002}, dipole oscillations~\cite{wolf_2010}, 
dissipative Rydberg atoms~\cite{ray_16}, etc. can well be described in the 
framework of Gutzwiller approximations. Also, it has been shown that GMFT can 
reasonably describe particle loss in the quantum Zeno regime of bosonic 
systems with long-range interaction~\cite{vidanovic_14}. Additionally, more 
recent experimental results of Mott-superfluid crossover in the presence of 
localized dissipation has been explained well within the 
GMFT\cite{tomita_17}.

In what follows, we use the GMFT for density matrices and Lindblad master 
equation to study the particle loss in the different correlated superfluid 
regimes. In the WCSF regime, we find that, for weak dissipation, the loss 
rate is proportional to the strength of the dissipation. However, for 
stronger dissipation, the systems protects itself from being dissipated, hence
the loss rate decreases, corroborating the commonly known Zeno effect in the 
superfluid phase when the onsite repulsive interactions are weak. 
The non-monotonic nature of the particle loss is in stark contrast 
to the superfluid regime when interaction is relatively strong. 
In particular, we see Zeno-anti-Zeno crossover as a function of dissipation 
strength. We furthermore show that, in the SCSF, the long-time dynamics differs 
significantly from the WCSF. These facts also translate into the fidelity 
obtained in these two regimes. Thus the dissipative dynamics in the 
two-dimensional Bose-Hubbard model may indicate the effect of interaction in 
the coherent many-body phases of this system.

The rest of the paper is organized as follows. In Sec.~\ref{model}, we
discuss the model and formalism in the presence of local dissipation. In
particular, we demonstrate single-site particle loss using density matrix
formalism. This is followed by Sec.~\ref{results}, where we discuss
particle loss in the different regimes of the onsite interaction of the
model Hamiltonian. We also present the long-time dynamics in those regimes.
We furthermore show the evolution of fidelity for different strengths of the 
dissipation. Finally, we conclude with a discussion on the possible 
future directions in Sec.~\ref{conclusions}.


\section{Model and formalism}\label{model}
 
We consider a homogeneous bosonic gas at zero temperature trapped in a
two-dimensional (2D) optical lattice potential. Employing the tight-binding
approximation, the statics and dynamics of the bosonic atoms can be well
described by the single lowest band Bose-Hubbard (BH) Hamiltonian with
nearest-neighbour (NN) hopping between the lattice sites, and local onsite
repulsive interactions $(U)$ between the atoms: 
\begin{eqnarray}
 \hat{H} = -J\sum_{\langle ll' \rangle}{\hat a}_l^\dagger {\hat a}_{l'}
 +\frac{U}{2}\sum_l \hat{n}_l(\hat{n}_l -1) - \sum_i \mu {\hat n}_l,
 \label{bhm}
\end{eqnarray}
where $\langle ll' \rangle$ refers to nearest-neighbours $l$ and $l'$, 
$J$ is the hopping strength between two NN sites, and $\mu$ is the chemical 
potential; ${\hat a}_l^\dagger ({\hat a}_l)$ are the bosonic 
creation (annihilation) operators with ${\hat n}_l = {\hat
a}_l^\dagger {\hat a}_l$ as the occupation number. 
Depending on the relative values of 
$J/U$, this model supports two distinct phases. For $J/U\ll 1 $, the system 
exhibits insulating phase, namely Mott-insulating (MI) phase with 
commensurate integer fillings and vanishing order parameter. 
In contrast, $J/U\gg1$ leads to superfluid (SF) phase with non-vanishing 
compressibility~\cite{fisher_89,jaksch_98,greiner_02,sansone_08}. As 
mentioned before, the superfluid medium can be further divided into
two regimes. Near the MI-SF boundary (at the onset of superfluidity), the 
superfluid states are strongly correlated with highly suppressed
superfluid stiffness, whereas $J/U\gg1$ corresponds to weakly-correlated 
superfluid with stiffness close to unity. Note that such correlated  phases 
can be distinguished by the non-perturbative renormalization group analysis 
as shown in Ref.~\onlinecite{rancon_11}.
Although distinction between these phases are beyond the scope of
Gutzwiller mean-field theory, however, it can estimate the superfluid 
stiffness \cite{krauth1992} which qualitatively agrees with the stiffness 
presented in Ref.~\onlinecite{rancon_11}. 

To obtain the ground state of BH model, we employ the mean-field
approximation by decomposing ${\hat a}_l = \phi_l + \delta {\hat a}_l$, 
where the order-parameter $\phi_l = \langle{\hat a}_l\rangle$. With 
this, Eq.~(\ref{bhm}) can further be simplified as a sum of 
single-site mean-field 
Hamiltonian $\hat H=\sum_{l,l'}\hat h^{\rm MF}_{l,l'}$ with
\begin{eqnarray}
 {\hat h}_{l,l'}^{\rm MF} &=& -J \big[(\phi_{l+1,l'}^{*}{\hat b}_{l,l'} -
 \phi_{l+1,l'}^{*}{\phi}_{l,l'}+ \phi_{l,l'-1}^{*}{\hat b}_{l,l'}\nonumber\\&& -
 \phi_{l,l'-1}^{*}{\phi}_{l,l'})+{\rm H.c.}\big]
 +\frac{U}{2}{\hat n}_{l,l'}({\hat n}_{l,l'} -1) - \mu{\hat n}_{l,l'},\nonumber\\
\label{sigH}
\end{eqnarray}
where $l$ and $l'$ are the lattice site indices along the $x$ and $y$
directions, respectively. In the Fock basis of occupation 
numbers, Eq.~({\ref{sigH}}) can eventually be diagonalised independently 
using the Gutzwiller variational ansatz
\begin{eqnarray}
|\Psi_{\rm GW}\rangle = \prod_{\otimes l,l'}|\psi\rangle_{l,l'} =
\prod_{\otimes l,l'}\sum_{n=0}^{N_{\rm b}}c_n^{(l,l')}|n\rangle_{l,l'},
\end{eqnarray}
where $|\psi\rangle_{l,l'}$ is the ground state of the lattice site with
index $(l,l')$, $|n\rangle_{l,l'}$ is the corresponding occupation number
basis, $c_n^{(l,l')}$s are the associated complex coefficients of
the expansion, satisfying $\sum_n|c_n^{(l,l')}|^2 = 1$, and $N_{\rm b}$ is 
the total number of basis states. Using these definitions and considerations,
the order parameter at the $(l,l')$th lattice site is given by

\begin{figure}
	\includegraphics[height=6.0cm,trim={0.5cm 2.5cm 2.5cm 0.25cm},clip]
	{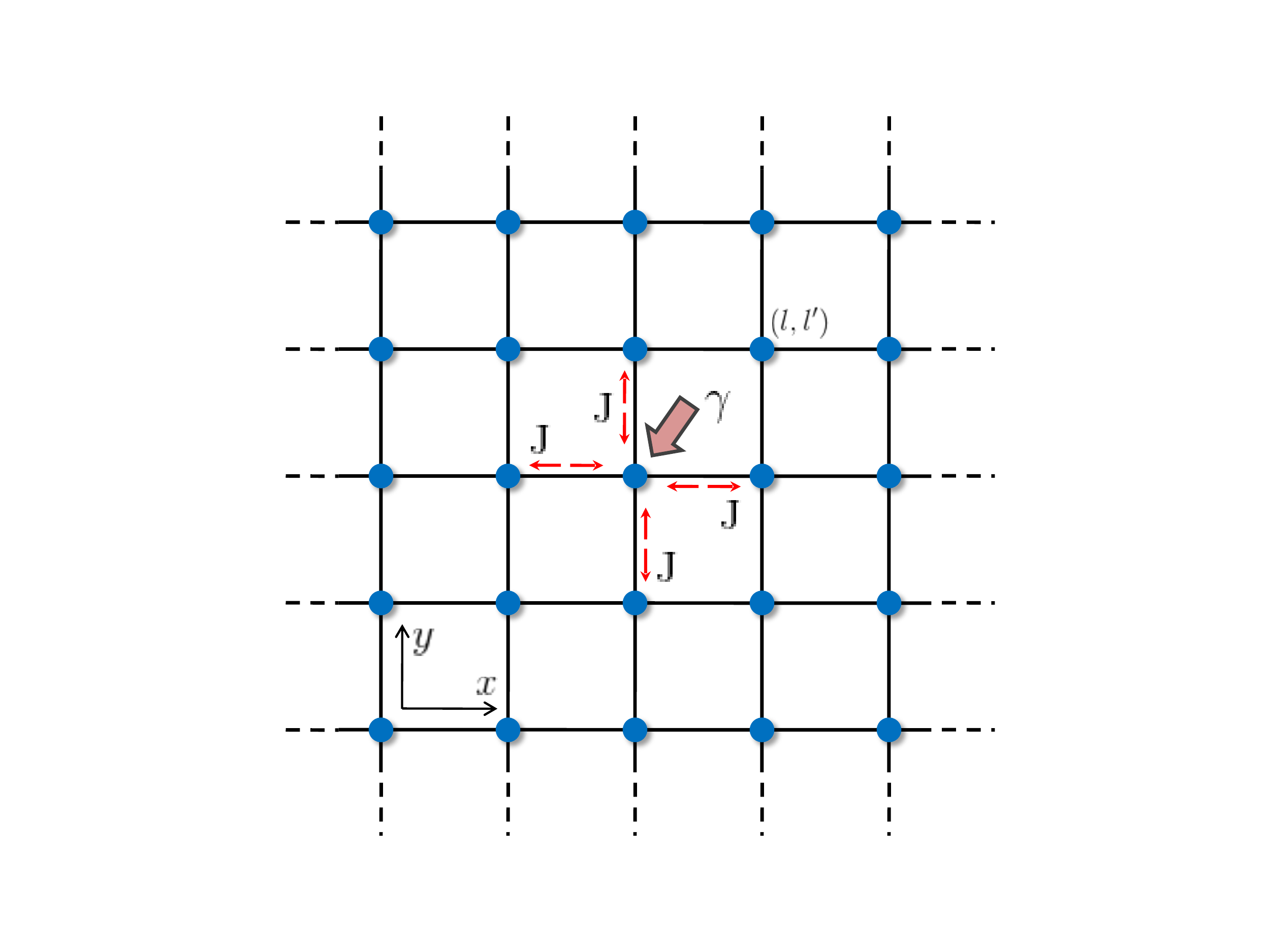}
	\caption{Schematic representation of a 2d lattice in which the 
             central lattice site is subjected to one-body dissipation with 
             strength $\gamma$. The red dashed arrows represent the nearest 
             neighbor tunneling with coupling strength $J$.
	        }
	\label{latticeg}
\end{figure}
\begin{eqnarray}
 \phi_{(l,l')} = \langle\Psi_{\rm GW}|{\hat b}_{l,l'}|\Psi_{\rm GW}\rangle
 = \sum_{n=1}^{N_{\rm b}}\sqrt{n}c_{n-1}^{*(l,l')}c_{n}^{(l,l')}
\end{eqnarray}
As mentioned earlier, $\phi_{(l,l')} =0$ in the MI phase, whereas it is
finite in the SF phase. Note that, in our GMFT analysis, the critical
values of hopping $J_c/U$, at which the SF order parameter starts to
develop, is obtained to be $0.04$ at $\mu/U=0.5$. Thus, within the
framework of GMFT, we consider $J/U\sim 0.04-0.05$ as strongly-correlated
SF regime, whereas $J/U\gtrsim0.1$ as weakly-correlated SF regime for
$\mu/U=0.5$. It is worth mentioning that in addition to equilibrium GMFT, the 
time-dependent GMFT has been successful in capturing the essential physics 
related to amplitude modes on interacting bosons in a lattice 
model~\cite{bissbort_11}.


\begin{figure*}
	\includegraphics[width=0.95\linewidth]{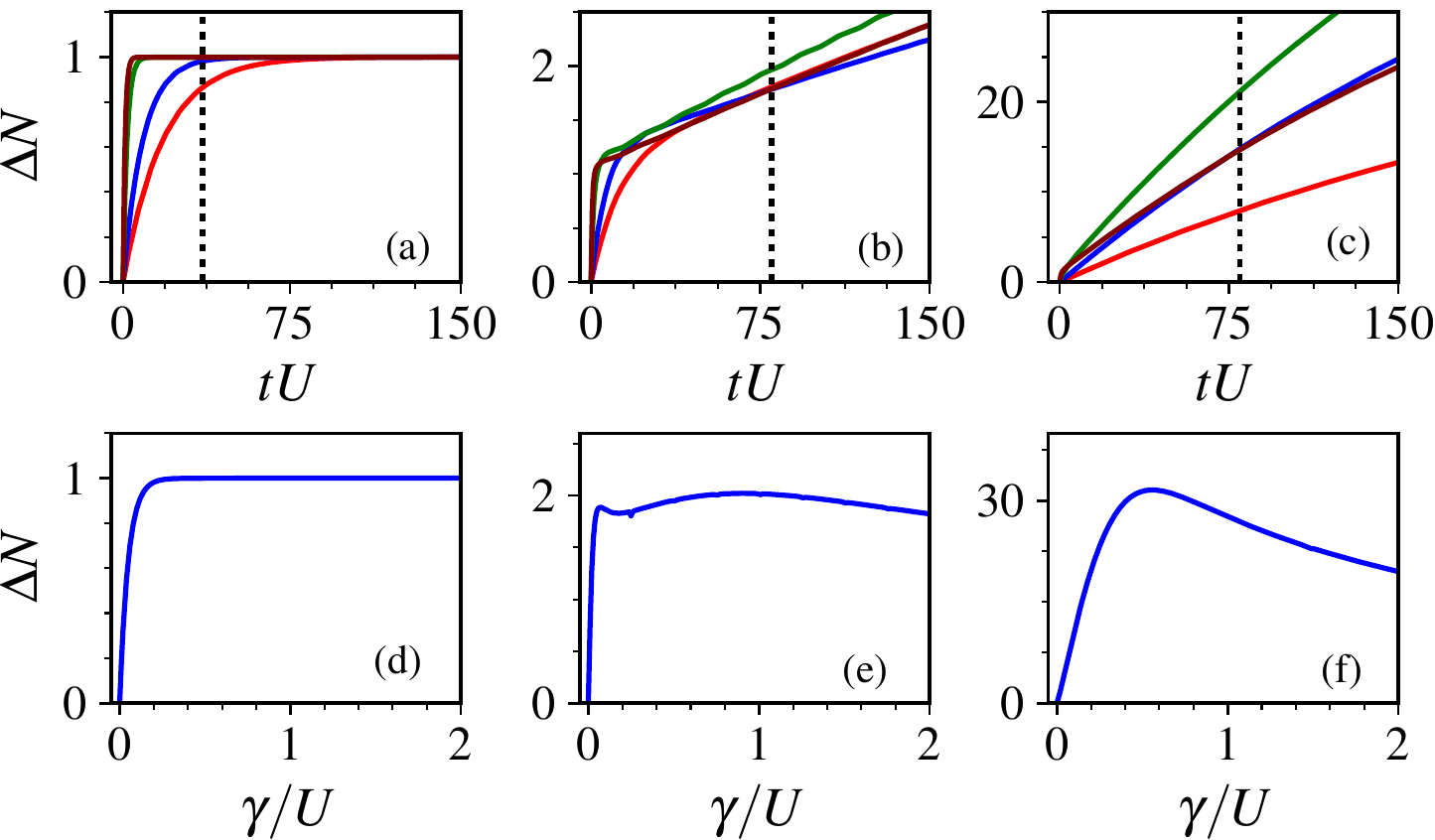}
	\caption{(a),(b),(c) Variation of $\Delta N$ with time for $J/U =
		0.02,0.05$ and $0.1$ respectively. In each plot: the red, green,
		blue and maroon lines denote $\gamma/U = 0.1,0.2,1.0$ and $2.0$
		respectively. (d),(e),(f) Variation of $\Delta N$ with $\gamma/U$
		for $tU=20,80,80$ marked with a dotted line in the corresponding
		adjoining figure.}
	\label{deltan}
\end{figure*}

\subsection{One-body loss}

Having obtained the many-body ground states for the BH model through GMFT, we 
wish to look at the effects of dissipation on these phases. We first 
introduce a local loss term that acts on a single site of the lattice 
system to remove particles. The loss can either be one-body, two-body or 
higher orders. Invoking Born-Markov
approximation and rotating-wave approximation~\cite{breuer_02}, the lossy 
dynamics of the BH model can thus be represented through the following 
Lindblad equation with local dissipation rate $\gamma$, which can
experimentally be tuned by varying the intensity of the applied 
beam, as ~\cite{barontini_13}
\begin{eqnarray}
\label{rhot}
 {\partial_t{\hat\rho}} = -i[{\hat H},{\hat \rho}] 
 + \gamma\big[{\hat a}_j{\hat \rho} {\hat a}_j^\dagger -
 \frac{1}{2}\{{\hat a}_j^\dagger {\hat a}_j,{\hat \rho}\}\big].
\end{eqnarray}
The first term on the right represents unitary dynamics of the BH model,
whereas the second term describes the non-unitary evolution. Applying the 
Gutzwiller approximation to the density operator 
$\rho = \prod_{\otimes
l,l'}\sum_{n,m}c_{nm}^{(l,l')}|n\rangle_{(l,l')}\langle m |_{(l,l')}$,
the evolution of the matrix elements of the density operator corresponding
to a lattice site is obtained to be
\begin{eqnarray}
 {\partial_t c_{nm}^{(l,l')}} &=& i\left\{J(\sqrt{n}c_{n-1,m}^{(l,l')}
 -\sqrt{m+1}c_{n,m+1}^{(l,l')})(\phi_{(l+1,l')}\right.\nonumber\\
 &+&\phi_{(l-1,l')}+\phi_{(l,l'+1)}+\phi_{(l,l'-1)})
 +J(\sqrt{n+1}c_{n+1,m}^{(l,l')}\nonumber\\
 &-&\sqrt{m}c_{n,m-1}^{l,l'})(\phi_{(l+1,l')}^*+\phi_{(l-1,l')}^*
 +\phi_{(l,l'+1)}^*
 \nonumber\\&+& \phi_{(l,l'-1)}^*)+(\mu(n-m)-\frac{U}{2}[n(n-1)-m(m-1)]
 \nonumber\\
 &+&\left.i\frac{\gamma}{2}(n+m))c_{n,m}^{(l,l')}\right\}
 +\gamma\sqrt{(n+1)(m+1)}c_{n+1,m+1}^{(l,l')}.\nonumber\\
 \label{cnm}
\end{eqnarray}

Clearly Eq.~(\ref{cnm}) is a coupled equation of multiple variables which
entails analytical solutions at some special values or cases 
which will be evident shortly. Eq.~(\ref{cnm}) also involves several 
parameters of the Hamiltonian. Thus, the dynamics is expected to depend on the 
interplay between those parameters, which may lead to interesting dissipative 
dynamics.

To solve Eq.~(\ref{cnm}) numerically, we use the fourth-order Runge-Kutta 
method with open boundary condition with the order parameter $\phi$ of the BH 
Hamiltonian being obtained by imaginary time propagation. Using the 
coefficients of expansion $c_{nm}$, one can obtain the total number of lost 
particles,
\begin{equation}
 \Delta N = N(t=0)-N(t) = \langle n \rangle(t=0) - \langle n \rangle(t),
\end{equation}
where $N(t)=\sum_{l,l'}\sum_n c_{nn}^{(l,l')}(t)|n|^2$ is total number of 
particles in the system.

We finally compute the time evolution of fidelity~\cite{sommer_2005} 
$F={\rm Tr}({\hat \rho}(t){\hat \rho}(0))$ as a measure of probability to 
recover the initial state at a given time, $t$. For the present work, 
we evaluate $F(t)$ for different strengths of the dissipation $\gamma$ and 
discuss its non-monotonic dependence on $\gamma$ as a manifestation of 
unconventional particle loss in the correlated superfluid regime. 

\section{Results and discussions}\label{results}
We consider a 2d lattice potential in which the central site is subjected 
to dissipation mediated through one-body loss as shown in
Fig.~\ref{latticeg}. Theoretically, the dynamics of the composite system is 
governed by the Lindblad equation in the Gutzwiller approximation following
Eq.~(\ref{cnm}). On the experimental front, such a setup is feasible in which 
the central site is coupled to a localized bath and the resulting dynamics is
monitored~\cite{barontini_13}. For the current study, we choose a
finite-sized $41\times41$ lattice system and $N_b=10$. We checked that this
system size is sufficient to reveal correct dynamics in the parameter
regime considered here. We then vary $J$ to obtain the initial homogeneous 
ground state in the three different 
regimes, namely ground states of (a) the deep SF, (b) the strongly-correlated 
SF, and (c) the Mott insulating regime.

\subsection{Particle loss}
The coupling between the system and bath is turned on at $t=0$,
which drives away the atoms from the central site, and a defect is thus
formed. The central site tends to lose particles, and eventually gets
refilled by the nearest neighbour tunneling processes. For $t>0$, the
dynamics includes hopping processes propagating through the system into the
lossy site which gives rise to an effective coupling between the NN sites.
This leads to the whole system getting affected as a result of localized
dissipation. The difference in the rate of loss of particles and the rate
of refilling leads to a variation in the
dissipative dynamics which forms the motivation of our present investigation. 
To quantify this behaviour, we compute the total number of particles lost 
from the system, and demonstrate that the intermediate and long-time 
behaviour for the three different phases is indeed different upon the 
application of a localized dissipation. Furthermore, we also compute the 
fidelity to substantiate our findings.

{\em Mott Phase--}
When $J/U \ll 1$, that is in the MI regime with
$\langle n \rangle(t=0) = 1$, the loss rate monotonically increases
with $\gamma$, which occurs due to randomization of phases in this 
regime~\cite{kaulakys_97,smerzi_02}. The corresponding plot is shown in 
Fig.~\ref{deltan}(a),(d). On varying $\gamma$, and choosing an intermediate
time $tU = 20$, we find $\Delta N$ to be increasing and gets saturated.
This can be understood from Eq.~(\ref{cnm}). Since the superfluid order 
parameter $\phi$ remains zero in the MI phase for all 
values of $\gamma$, Eq.~(\ref{cnm}) for $n=m$ leads to 
\begin{align}
c_{nn}^{(l,l')}(t)=c_{n,n}^{(l,l')}(0)e^{-\gamma t},
\label{mottcnm}
\end{align}
where we have used $c_{n+1,n+1}=0$ due to the Mott nature of the
wavefunction. This gives $\Delta N=1-e^{-2\gamma t}$ and
validates the result of Fig.~\ref{deltan} (a),(d).

\begin{figure}
 \includegraphics[width=0.99\linewidth]{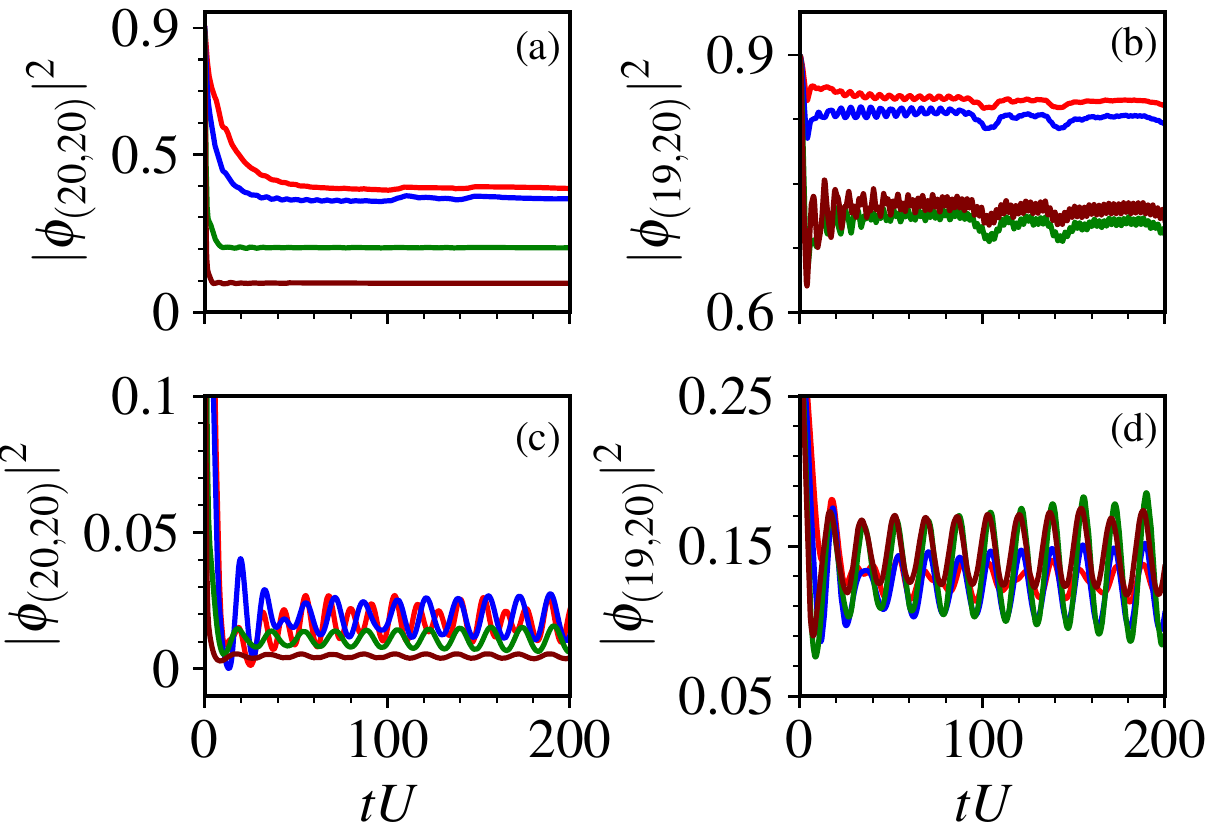}
 \caption{Evolution of order parameter ($|\phi(t)|^2$) deep into the
          superfluid regime for (a) central site with local dissipation and (b)
          nearest four site of the central site. Since  $|\phi_{l,l+1}|^2$ is 
          same for all sites nearest to $\phi_{l,l}$, we only plot one of 
          them for various strengths of the dissipation. Evidently, the Zeno 
          effect is reflected in the order parameters of nearest sites with 
          the central site coupled to local dissipation. In each 
          plot: the red, green, blue and maroon lines 
          denote $\gamma/U = 0.1,0.2,1.0$ and $2.0$ respectively. 
          (c),(d) Same as (a),(b) for strongly-correlated superfluid regime.}
 \label{orderp}
\end{figure}

{\em Deep SF Phase--} In the deep SF regime, we can ignore the term
involving $U$ in Eq.~(\ref{cnm}). Then for fixed $\mu$, the competition
between $J$ and $\gamma$ leads to non-monotonic particle loss as a function
$\gamma$. Fig.~\ref{deltan}(c),(f) reflect this
behavior, where the loss rate increases linearly with $\gamma$ for small
$\gamma$. This can be qualitatively understood from Eq.~(\ref{cnm}) and
using some perturbative argument.  We first set $\gamma=0$ to find
$c_{nm}(t)$. This corresponds to equilibrium dynamics of a closed BH model.
In this limit, $c_{nn}(t)$ turns out to be constant as also evident from
Eq.~(\ref{rhot}). This is essentially because the order parameter $\phi$
is real, and $\rho$ is symmetric. Albeit, $\gamma \neq 0$ does not render 
$c_{nn}(t)$ to be constant.
Using this solution as an initial solution for finite but
very small $\gamma$ in Eq.~(\ref{cnm}), we obtain $c_{nn}(t)\sim
\alpha_0+\gamma t$, where $\alpha_0$ is a constant function of $J$. Note
that, the order parameter $\phi$ of the central site and its nearest sites
have been considered to be constant for some accessible time domain as seen
in Fig.~\ref{orderp}.  Note also, the approximate solution of $c_{nn}(t)$
is valid only up to some intermediate time. Thus for fixed time we 
obtain $\Delta N\sim\gamma$ for small $\gamma$ and it qualitatively agrees 
with the perturbative analysis of one-dimensional hard-core bosons in optical
lattices\cite{ripoll_09}. 

On the other hand, strong dissipation $\gamma \gg J$ leads to quantum Zeno
effect which describes the suppression of dynamics due to the application
of a continuous projective measurement. In this limit, the system ceases to
get affected by attaining a dark
state~\cite{ripoll_09,bardyn_13,fischer_01}. Consequently, the loss of
atoms will be suppressed as evident from Fig.~\ref{deltan}(c),f. To
understand this, we neglect terms involving $J$ since $\gamma\gg J$.
Then $c_{nn}(t)$ can be approximated as $c_{nn}(t)\sim e^{-\gamma t}$,
considering minimal contribution from nearest sites. Note that for stronger
dissipation, the order parameter reaches to steady states much faster than
weak dissipation. However, the Zeno behavior is reflected only in the order
parameter of the nearest sites. Fig.~\ref{orderp}(b) clearly shows the
non-monotonic behaviour of $|\phi_{l,l+1}(t)|^2$ for strong dissipation.

{\em Close to SF-Mott boundary--} A dramatic change in the dynamical
behaviour occurs for the phase close to the SF-MI transition, that is when $J/U = 0.05$. This phase also referred to as the strongly-correlated 
superfluid phase where the condensate density is suppressed, exhibiting a 
different behavior upon the application of a localized dissipation.  It is 
evident from Fig.~\ref{deltan}(b),(e)  that, for intermediate time, 
$\Delta N$ initially increases with the increase of $\gamma$ as before. 
However, for moderate range of $\gamma$, we see decaying and increasing trend 
of $\Delta N$. This is in contrast to the particle loss both in the MI and 
deep SF phase. In addition, for stronger dissipation, $\Delta N$ decays 
notably slow as compared to the deep SF regime. In a nutshell, we find 
Zeno-anti-Zeno behavior at the onset of superfluidity (near the MI-SF phase 
boundary). Indeed, here the dynamics is governed by the combination of all 
parameters $J,U,\gamma$. As a result, the evolution of order parameters in 
the central site and its nearest site differs significantly than their 
dynamics deep into the superfluid (see Fig.~\ref{orderp} (c), (d)), hence 
the particle loss.

To this end, we comment on the dynamics away from $\mu=0.05$. 
For fixed $J/U = 0.05$, the Zeno-anti-Zeno
regime is retained with slight variation of $\mu$. However, such regime turns
out to be very narrow and asymmetric with respect to $\mu$. Indeed, this is 
attributed to the asymmetric SF-MI phase boundary. 
\begin{figure}
	\includegraphics[width=8.0cm]{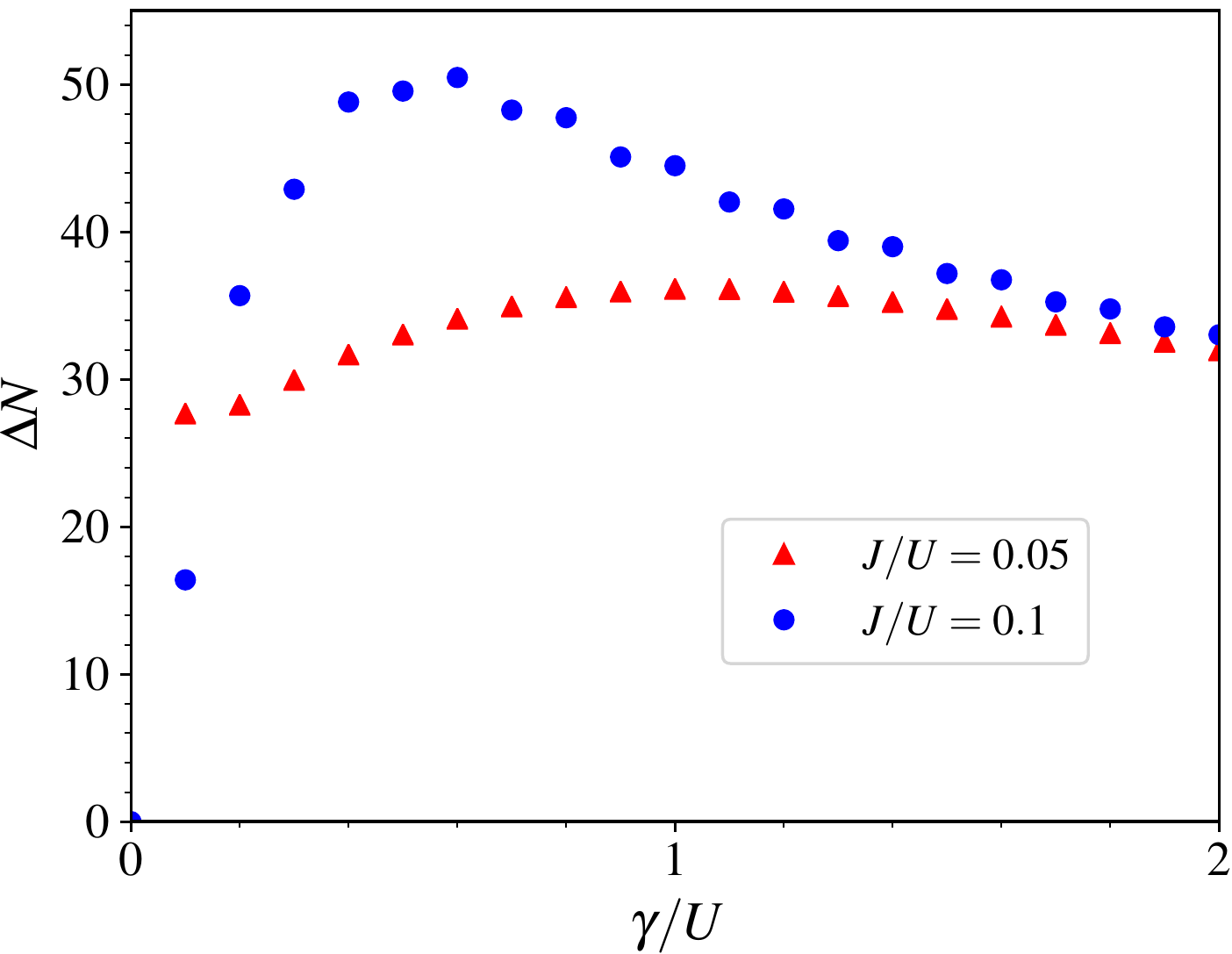}
	\caption{ Variation of particle loss $\Delta N$ with dissipation
		$\gamma$ at time $tU=200$. Clearly, the long-time dynamics mimics the
		intermediate-time dynamics of particle loss in the weakly correlated
		superfluid regime. In contrast, particle loss in the
		strongly-correlated superfluid regime is almost negligible. For 
        better visualisation, 
        the particle loss in SCSF has been enhanced ten times in
		this plot.}
	\label{fig:long-time}
\end{figure}

\subsection{Long-time dynamics}
In this section, we discuss the long-time dynamics of particle loss for
different strengths of dissipation $\gamma$. In the weakly-correlated
superfluid regime, the long-time dynamics turns out to be similar to the
dynamics of intermediate time. This is apparent in
Fig.~\ref{fig:long-time} where the particle loss initially increases with
$\gamma$ and acquires some maximum value at some particular $\gamma$, and
finally it decreases as a consequence of Zeno effect. This resembles to the
particle loss at $tU=80$ as seen from Fig.~\ref{deltan}(d). In contrast, 
the Zeno-anti Zeno kind of behaviour is not prominent in the long-time 
dynamics of the strongly-correlated regime [see Fig.~\ref{deltan} (e)]. 
In fact, the loss is almost same for all range of $\gamma$. The reason for 
such behavior can again be attributed to the randomization of phases close to 
the phase boundary similar to the Mott phase.
As the oscillation amplitudes of the order parameter increase in the 
long-time limit (cf. Fig~{\ref{orderp}} (c),(d)), we can approximate the 
average values of the order parameter to be zero.
This leads to $\Delta N\sim \mathcal{O}(1)$ 
similar to the Mott phase.
Thus this distinction of particle loss in the long-time limit is in
conjunction with our earlier observations that the dissipative dynamics in
bosonic systems strongly depends on the local interaction.

\begin{figure}
	\includegraphics[width=8.5cm]{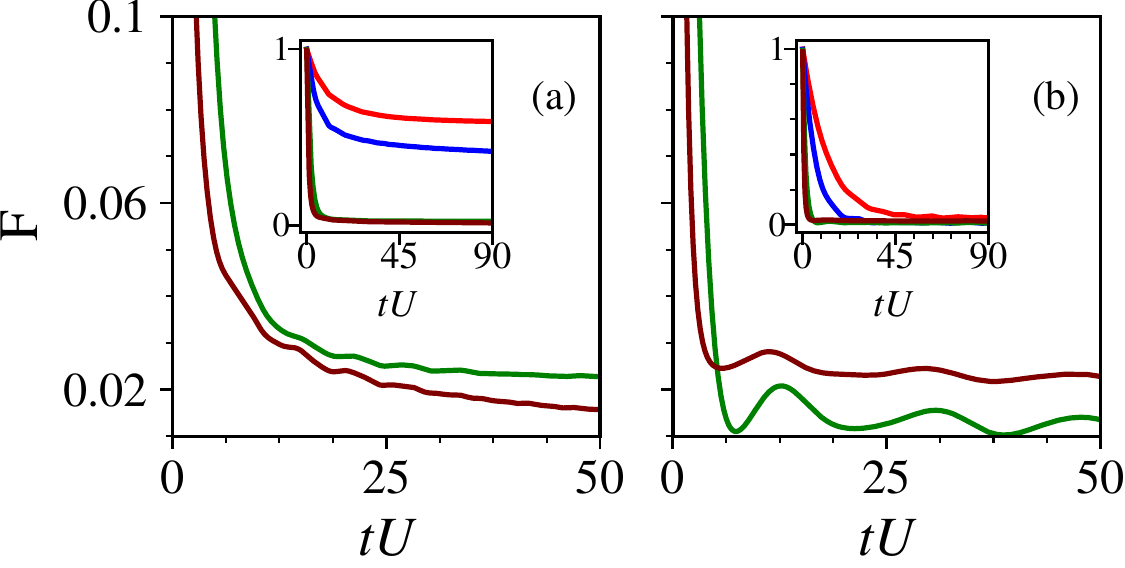}
	\caption{Variation of fidelity in the (a)weakly, and (b)strongly 
		correlated regime. In each plot: the red, green,
		blue and maroon lines denote $\gamma/U = 0.1,0.2,1.0$ and
		$2.0$ respectively.}
	\label{fd}
\end{figure}
\subsection{Fidelity}
To substantiate the results obtained in the preceding sections, we discuss
the evolution of fidelity both in the WCSF and SCSF regimes.
Fig.~\ref{fd}(a)-(b) illustrate the fidelity in both the superfluid regimes
for various $\gamma$. Clearly, $F(t)$ decays exponentially as a function
    of $t$ and finally saturates after an accessible time, corroborating
    the non-equilibrium steady state behavior of the
    system\cite{gorin2004}. However, the degree of decay depends on the
    strength of the dissipation. For stronger dissipation, $F(t)$ saturates 
    even before relatively weaker
    dissipation in the SCSF regime (cf. Fig.~\ref{fd}(b)). This is in
    contrast to the case of WCSF where no such behavior is observed (cf.
    Fig.~\ref{fd}(a)). This is again attributed to the different 
    dynamical behavior of the particle loss in different regime of the 
    SF-MI transition. Since it is experimentally feasible to measure fidelity 
    in open quantum dynamics, it can be used as a probe to identify different 
    interacting regimes of the Bose-Hubbard model
    ~\cite{wimberger_16,talukdar_10,ullah_11,arimondo_11,bederson_99}.


\section{Conclusions}\label{conclusions}
In conclusion, we present the dynamics of a two-dimensional Bose-Hubbard
model in the presence of a local dissipation. We find that the particle
loss in this setting strongly depends on the on-site bosonic interactions.
While the particle loss in the Mott phase has monotonic behavior as a
function of dissipation, it significantly differs in the superfluid regimes
with non-monotonic dependence on the strength of the dissipation. In
addition, we find notably different dissipative dynamics in the strongly
and weakly correlated superfluid regimes. Such differences are 
reflected in the fidelity and long-time dynamics.

Although the results reported here are explained qualitatively within the
framework of Gutzwiller mean-field theory for density matrices and Lindblad
master equation, it would be desirable in future to  verify it using more
sophisticated tools like truncated Wigner approximations~\cite{kordas_15},
quantum trajectories method~\cite{daley_14}
or by constructing
density matrix using exact diagonalization techniques, alleviating the 
limitations of GMFT. A natural extension of the study presented here would
be to explore dissipative dynamics in disordered systems, and in particular, 
the Bose glass phase. 
 
\begin{acknowledgments}
We thank J. Marino, A. Lazarides and S. Diehl for useful discussions. AR
thanks R. Bai, S. Bandopadhyay, K. Suthar, S. Pal, and D. Angom for
insightful discussions on GMFT.

\end{acknowledgments}

\bibliography{refs}{}

\end{document}